\documentclass[aps,pre,twocolumn,superscriptaddress,showpacs,floatfix]{revtex4-1}

\usepackage{graphicx}
\usepackage{amssymb,amsmath,amsbsy,amsfonts,bm}
\usepackage{epsfig}
\usepackage{subfigure}
\usepackage{multirow}
\usepackage{tikz}
\usepackage{hyperref}
\usepackage{gensymb}
\usepackage{threeparttable}
\usepackage{csquotes}
\usepackage{braket}
\usepackage{cleveref}
\usepackage[percent]{overpic}
\usepackage{color}

\definecolor{darklavender}{rgb}{0.45, 0.31, 0.59}
\definecolor{amethyst}{rgb}{0.6, 0.4, 0.8}
\definecolor{paulcolour}{rgb}{0.78, 0.082, 0.52}

\newcommand{\ofr}{(\mathbf{r})}
\newcommand{\sect}{Sec.}
\newcommand{\ndens}{\rho}
\newcommand{\imag}{\text{\textbf{{\textit{i}}}}}

\newlength{\myl}%
\newcommand{\dif}{\mathrm{d}}%
\newcommand{\nofr}{\mathbf{\hat{n}}\ofr}%
\newcommand{\Sofr}{S\ofr}%
\newcommand{\Tofr}{T\ofr}%

\newcommand{\SUM}[2]{{\setlength{\myl}{\widthof{$\displaystyle\sum_{#1}^{#2}$}*\real{0.5}-\widthof{$\displaystyle\sum$}*\real{0.5}}\sum_{#1}^{#2}\;\hspace{-\the\myl}}}
\newcommand{\INT}[3]{\settowidth{\myl}{$\displaystyle\int_{#1}^{#2}$}{\int_{#1}^{#2}\;\;\;\hspace{-\the\myl}\dif #3}\,}
\newcommand{\uu}{\mathbf{\hat{u}}}%
\newcommand{\INTOII}{\INT{S^{2}}{}{\Omega}}%
\newcommand{\ff}[3]{f_{#3}^{(#1\mathrm{d})}}%
\newcommand{\TT}[3]{\mathcal{P}_{#3}^{(#1\mathrm{d})}}%

\usepackage[normalem]{ulem}%
\newcommand{\New}[1]{\textcolor{green}{#1}}%

\begin{document}
\title{Topological fine structure of smectic grain boundaries and tetratic disclination lines within three-dimensional smectic liquid crystals}

\author{Paul A. Monderkamp}
\affiliation{Institut f\"ur Theoretische Physik II: Weiche Materie, Heinrich-Heine-Universit\"at D\"usseldorf, 40225 D\"usseldorf, Germany}

\author{Ren\'e Wittmann}
\email{Rene.Wittmann@hhu.de}
\affiliation{Institut f\"ur Theoretische Physik II: Weiche Materie, Heinrich-Heine-Universit\"at D\"usseldorf, 40225 D\"usseldorf, Germany}

\author{Michael te Vrugt}
\affiliation{Institut f\"ur Theoretische Physik, Center for Soft Nanoscience, Westf\"alische Wilhelms-Universit\"at M\"unster, 48149 M\"unster, Germany}

\author{Axel Voigt}
\affiliation{Institut für Wissenschaftliches Rechnen, Technische Universität Dresden, 01062 Dresden, Germany}

\author{Raphael Wittkowski}
\affiliation{Institut f\"ur Theoretische Physik, Center for Soft Nanoscience, Westf\"alische Wilhelms-Universit\"at M\"unster, 48149 M\"unster, Germany}

\author{Hartmut L\"owen}
\affiliation{Institut f\"ur Theoretische Physik II: Weiche Materie, Heinrich-Heine-Universit\"at D\"usseldorf, 40225 D\"usseldorf, Germany}

\begin{abstract}
Observing and characterizing the complex ordering phenomena of liquid crystals subjected to external constraints constitutes an ongoing challenge for chemists and physicists alike.
To elucidate the delicate balance appearing when the intrinsic positional order of smectic liquid crystals comes into play, we perform Monte-Carlo simulations of rod-like particles in a range of cavities with a cylindrical symmetry.
Based on recent insights into the topology of smectic orientational grain boundaries in two dimensions, we analyze the emerging three-dimensional defect structures from the perspective of tetratic symmetry.
Using an appropriate three-dimensional tetratic order parameter constructed from the Steinhardt order parameters, we show that those grain boundaries can be interpreted as a pair of tetratic disclination lines that are located on the edges of the nematic domain boundary. Thereby, we shed light on the fine structure of grain boundaries in three-dimensional confined smectics.
\end{abstract} 

\maketitle
\section{\label{sec_introduction}Introduction}
Omnipresent throughout a vast range of chemical and physical systems \cite{lu2020meron,shinjo2000magnetic,phatak2012direct,tan2016topology,sorokin2017first,toptherdef,topcosmdomains,topdeftwocompAP,superconduc,ling2014photonic,guilian2016phononic,wang2020topological}, topological defects play a central role in characterizing collective ordering phenomena.
As the go-to systems for investigating these ordering phenomena,
liquid crystals have been enjoying continuous attention within the physical chemistry and chemical physics
community over the decades and remain an active field of research relevant 
in a variety of different applications. 
The most prominent examples are
colloidal systems \cite{kuijk2012phase,annulus,CollLQinSqConf}, various forms of passive and active chemical molecules \cite{tan2019topological,ndlec1997self,schaller2010polar,giomi2015geometry,decamp2015orientational,hamley2010liquid} and also living and artificial microscopic systems, such as swarms of bacteria \cite{beppu2017geometry,sokolov2007concentration,wensink2012meso,dunkel2013fluid}, bacterial DNA \cite{reich1994liquid,marchetti2013hydrodynamics} and viral colonies \cite{fowler2001tobacco},
that all exhibit liquid crystal mesophases and topological defects. 
Topological analysis even provides a tool for insight into the behavior of macroscopic systems, such as gracefully moving flocks of birds, often extending dozens of meters in the sky, as well as the collective behavior in shoals of fish, where local coherent swimming is a vital tool in the evasion of predators \cite{abaid2010fish,hall1986predator,ramaswamy2010mechanics,koch2011collective}. 

\begin{figure}[t]
\begin{center}
\includegraphics[width=0.8\linewidth]{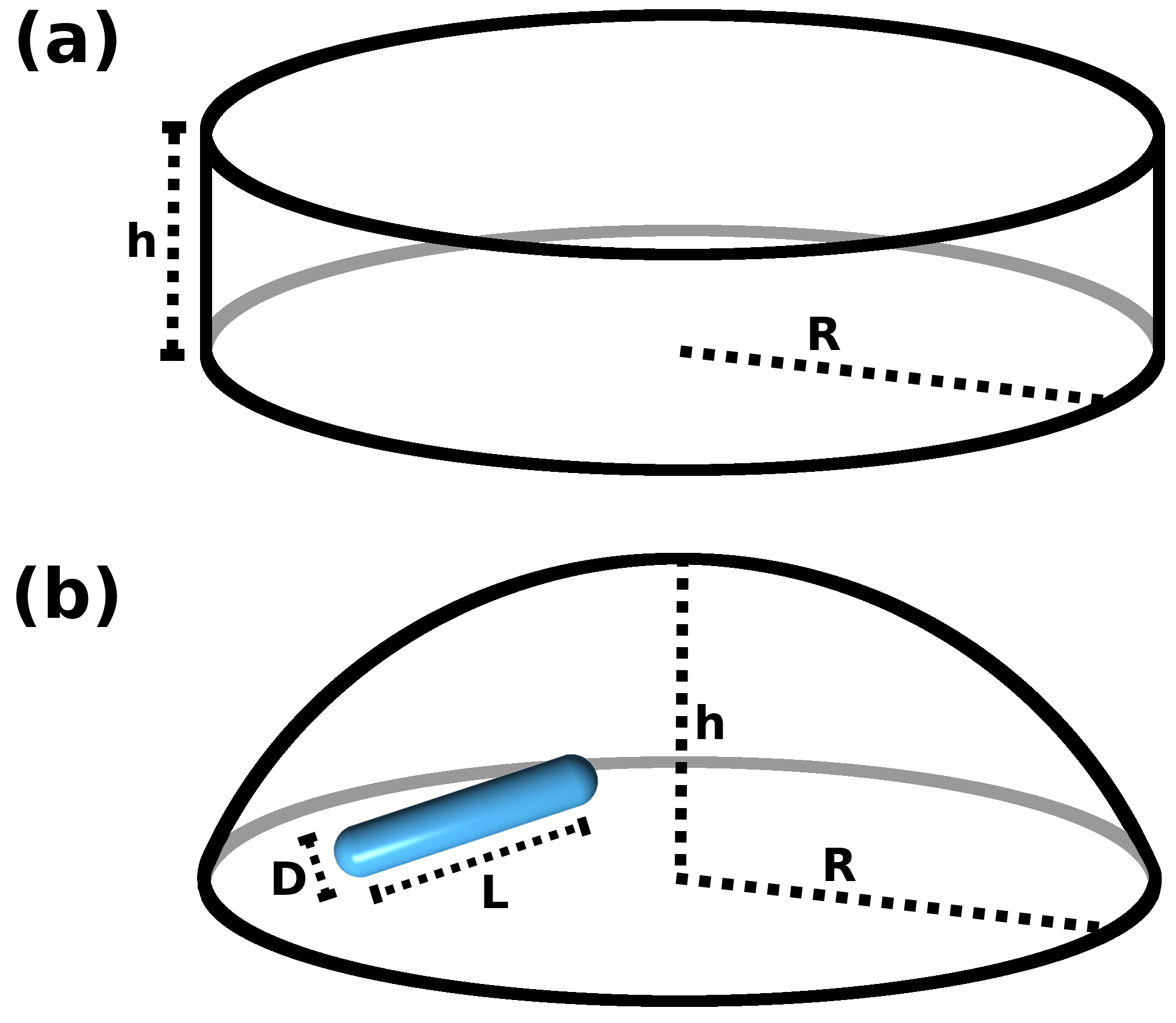}
\caption{\label{fig_concept}Schematic depiction of the confinement considered in our simulation of three-dimensional liquid crystals. We simulate the liquid crystals by a system of $N$ spherocylinders with length $L$ and diameter $D$ in confinement to suppress the bulk symmetry of the fluid in order to observe topological defects. \textbf{(a):} Cylindrical cavity with radius $R$ and height $h$. \textbf{(b):} Spherical cap cavity with radius $R$ and height $h$. In our simulations, we consider $R=4L$ and small heights $h$ up to $4.5L$.}
\end{center}
\end{figure}

The most prominent type of ordering, which is typically found in liquid crystals, is orientational (nematic) ordering, where the characteristically shaped subunits, i.e, molecules or colloidal particles in close proximity, show a tendency to align. If this preferred order gets frustrated, e.g., by confinement to a finite container \cite{lavrentovich1986phase,lavrentovich1998topological,real_defch,dammone2012,VirNemConfGeom,trukhina2008,varga2014hard,Jackson2017,majumdar2016multistable,rectangularConfinement,garlea2016finite,exp_studies,yaochen2020,basurto2020,yao2018topological,yao2021defect,dijkstra2001wetting}, constraining on a surface \cite{dzubiella2000topological,allahyarov2017,keber2014topology,Lubenskyetal_JPII_1992,Kralj_SM_2011,Nitschke_PRSA_2018,Nestler_SM_2020,Nitschke_PRSA_2020} or insertion of obstacles \cite{loewe2021passive,poulindirect,ruhwandl1997monte,andrienko2001computer,stabAndRew,ilnytskyi2014topological,schlotthauer2017,coll_transp,stark2001physics,kil2020}, topological defects emerge, which are discontinuities in the ordered structures that can display particle-like properties themselves \cite{vromans2016orientational,harth2020topological,hydroOfDef,tan2019topological,toptherdef}. 

In liquid crystals which feature exclusively orientational order of the fluid particles, so-called nematics, 
the commonly observed stable defects are singular points in a two-dimensional (2d) plane and curves in three-dimensional (3d) space that are either closed or end on system boundaries \cite{PhysRevA.43.6875}. 
The defect strength in 2d nematic liquid crystals is characterized by a so-called topological charge, which obeys an additive conservation law in analogy to the electric charge. This topological charge is determined by the total change of the preferred local orientation of the fluid on a contour around the defects
\cite{nem_defects,kleman1989defects}.
At higher densities and/or low temperatures, certain liquid crystals form the so-called smectic phase, which additionally displays positional order.
Traditionally, the study of defects in smectic liquid crystals is mainly concerned with the purely positional defects, such as edge dislocations \cite{EdgeDisloc,SmEdgeDislogShear,NanStrucEdgeDisloc,kamien2016} or complex structures like focal conics \cite{bramble2007,FocalDomains,liarte2015}.
However in situations, where frustration of nematic ordering takes a pronounced role, the consideration of the topology of the local orientations has proven insightful.

Smectic liquid crystals have a strong intrinsic tendency to maintain a uniform layering. The discontinuities in the order thus preferably appear as grain boundaries separating different domains within the fluid, which have a linear shape in two and a planar shape in three dimensions \cite{dozov1994quantized,kleman2000grain}. These grain boundaries are especially pronounced, when the fluid is confined to a container, where the local preferred orientation in the system depends heavily on the position. A convenient way to investigate these is therefore consideration in finite cavities.
For 2d colloidal liquid crystals, in particular, we have previously shown that the consideration of those domain boundaries as coexisting nematic and tetratic charges yields insight into the orientational topology of smectics \cite{monderkamp2021topology,annulus} (for a comprehensive summary see \sect~\ref{sec_topcharge}). 
The role of the orientations within smectic liquid crystals remains to be further understood. This concerns in particular the analysis of three-dimensional systems.

To shed more light on this issue, we present a range of simulation results for confined smectic liquid crystals in three dimensions, where the confinement causes frustration of the bulk symmetry and induces the formation of topological defects.
As elaborated in \sect~\ref{sec_topcharge}, the investigation of orientational topology in three dimensions is more involved, since 3d topological charges do not adhere to an additive charge conservation like their 2d counterparts \cite{nem_defects,afghah2018visualising,exp_studies}. However, under specific circumstances, this additive charge conservation is recovered.
By elaborating on the analogy to the 2d case, we explain the effects of the introduction of the third dimension and investigate the conditions for additive topological charge conservation.

This article is structured as follows: In \sect~\ref{sec_methods}, we present our methodology. Section~\ref{sec_simulation} elaborates on the simulation protocol, where we introduce the order parameters, used to characterize the simulation results, in \sect~\ref{sec_ordpar}. In \sect~\ref{sec_topcharge}, we explain the analysis of the topological charge and the details of the charge conservation. In \sect~\ref{sec_results}, we present our results, by first evaluating in \sect~\ref{sec_res_OP} the order parameters to detect and classify the emerging defects. Then we characterize the layer structure and orientations within the confinements in \sect~\ref{sec_conf_sm_structure}. We discuss the implications on the topological charge in \sect~\ref{sec_res_top_charge}. 
Lastly, we conclude in \sect~\ref{sec_conclusion}.

\section{\label{sec_methods}Methods}
\subsection{\label{sec_simulation}Simulations}
We perform canonical Monte Carlo (MC) simulations of systems of hard rods confined to 3d cavities, see Fig.~\ref{fig_concept}. Specifically, we consider soft walls in the shape of cylinders $\{ \mathbf{r} \in \mathbb{R}^3 |r_x^2 + r_y^2 \leq R^2,
0 \leq r_z \leq h \}$ and spherical caps, resembling a drop-like shape, $\{ \mathbf{r} \in \mathbb{R}^3 | r_x^2 + r_y^2 + r_z^2 < R^2,
r_z > R-h \}$, both of radius $R$ and height $h$. 
The rods are modeled as spherocylinders with aspect ratio $p = L/D = 5$, with core length $L$ and diameter $D$. 
Note that in two dimensions, one would require significantly longer rods to observe stable smectic structures.

The pair potential for the particle-particle interaction is given by the standard hard-core repulsion \cite{overl}
\begin{align}
U(\mathbf{r}_i,\mathbf{r}_j,\mathbf{\hat{u}}_i,\mathbf{\hat{u}}_j) = 
\begin{cases}
\infty & \text{ for } d_{i,j} \leq D, \\
0 & \text{ for } d_{i,j} > D
\end{cases}
\end{align}
for spherocylinders with 
\begin{equation}
d_{i,j} = \min_{\left | \alpha,\beta \right | < \frac{L}{2}} \left \| \mathbf{r}_i 
+ \alpha \mathbf{\hat{u}}_i - ( \mathbf{r}_j + \beta \mathbf{\hat{u}}_j ) \right \|,
\end{equation}
where $\mathbf{r}_k$ and $\mathbf{\hat{u}}_k$ are the position and normalized orientation of the $k$-th rod, respectively.
The convexity of the confining cavities enables us to specify a wall-particle interaction potential $V(x)$ by modeling the rods as two virtual point-like particles at $\mathbf{r}_\pm = \mathbf{r}_k \pm (L/2) \mathbf{u}_k$. The interaction potential is then given by
\begin{align}
\label{eq_wallpot}
V(x) =
\begin{cases}
\Phi(x_{0}) +\Phi'(x_{0})(x-x_0) & \text{ for } x \leq x_{0}, \\
\Phi(x) & \text{ for } x_{0} > x,
\end{cases}
\end{align}
where $\left | x \right |$ denotes the minimal perpendicular distance from either
of the two points to the wall and $x>0$ corresponds to the inside of the cavity. The cut-off point, below which $V(x)$ is linear, is chosen as $x_0=0.5D$. Moreover, $\Phi(x)$ is the canonical Weeks-Chandler-Andersen-potential
\begin{equation}
\Phi(x) =
\begin{cases}
4\epsilon \left [ \left ( \frac{D}{x} \right )^{12} - \left ( \frac{D}{x} \right )^{6} \right ] + \epsilon & \text{ for } x \leq 2^{\frac{1}{6}} D \\
0 & \text{ for } x > 2^{\frac{1}{6}} D, 
\end{cases}
\end{equation}
with $\epsilon = 10 k_B T$ (with the Boltzmann constant $k_B$ and the temperature $T$) \cite{WCA}, mimicking nearly hard walls. 
In what follows, we consider confinements with fixed footprint radius $R = 4L$ and different heights $h \in (0,4.5L]$.

To create the smectic structures in our 3d cavities,
we follow a compression protocol, where we initialize the system at a low volume fraction $\eta_0 = 5 \times 10^{-3}$, compress with a rate of $\Delta \eta_1= 2.45 \times 10^{-7}$ per MC cycle to a bulk-isotropic volume fraction $\eta_1 = 0.25$ 
and then, in a second stage, with a rate $\Delta \eta_2= 5.4 \times 10^{-8}$ per MC cycle to a bulk-smectic volume fraction $\eta_2 = 5.2$. Here, the volume fraction $\eta$ is defined as $\eta = NV_\text{hsc}/V_\text{cav}$, with the particle number $N$, the volume of a hard spherocylinder $V_\text{hsc}$ and the total volume of the confining cavity $V_\text{cav}$. 
Since we fix in each simulation run the particle geometry, the shape and size of the confinement and the final volume fraction, the particle number $N$ is a variable that gets adjusted accordingly. The values of $N$ we investigate, determined by the parameters above,
typically lie between $\approx 200$ for extremely shallow cavities and $\approx 3300$ for the tallest cavities. 

\subsection{\label{sec_ordpar}Order parameters}
We examine the structure of the confined fluid with the help of two orientational order parameters. The first one is the standard nematic order parameter $S$, associated with orientational ordering of uniaxial particles, which corresponds to the largest eigenvalue of the nematic tensor $\mathcal{Q}$ \cite{te2020relations,SoftMPhys,PhLiqCrys}. 
To numerically generate the scalar field $S(\mathbf{r})$, we sample the nematic tensor within a spherical subsystem of radius $2.5D$ around each point $\mathbf{r}$ as
\begin{align}
\mathcal{Q}(\mathbf{r}) = \left \langle \frac{3}{2} \uu_k \otimes \uu_k - \frac{1}{2} \mathbb{I}_3 \right \rangle_{B_{2.5D}(\mathbf{r})}\,.
\label{eq_Qsample}%
\end{align}
Here, the brackets denote an average over all $N_B$ particles contained within the ball $B_{2.5D}(\mathbf{r})$
with the individual orientations $\uu_k=(\sin\theta_k\cos\phi_k,\sin\theta_k\sin\phi_k,\cos\theta_k)^\mathrm{T}$ in spherical coordinates for $k\in\{1,\ldots,N_B\}$, where $\theta$ and $\phi$ are the angles to the $z$- and $x$-axes, respectively, and the 3d unit matrix $\mathbb{I}_3$.
$S(\mathbf{r})$ denotes the largest eigenvalue of $\mathcal{Q}(\mathbf{r})$.
The mean local orientation $\mathbf{\hat{n}}(\mathbf{r}) = \mathbf{n}(\mathbf{r}) / \left \| \mathbf{n}(\mathbf{r}) \right \|$ of the rods, i.e., the nematic director, can be computed by  normalizing the eigenvector $\mathbf{n}(\mathbf{r})$  associated with $S(\mathbf{r})$, where $\left \| \cdot \right \|$ is the Euclidean norm.

As will be discussed later, the favorable nematic bulk symmetry of orientational ordering is broken when the fluid is confined to a cavity. In two dimensions, the topological fine structure of the spatially extended defect lines in the director field $\mathbf{\hat{n}}(\mathbf{r})$ can be investigated using a scalar tetratic order-parameter field, which can be defined as 
\begin{equation}
\begin{split}
&T_\text{(2d)}(\mathbf{r}) =| \langle \exp(\imag4\phi_k)\rangle_{B_{2.5D}(\mathbf{r})} | 
\end{split}
\end{equation}
for a 2d subsystem with radius $2.5D$, with the imaginary unit \imag~and the 2d polar angle of the $k$-th particle $\phi_k$ \cite{sitta2018liquid,sanchez2015concentric,zhao2007nematic}.
Note that this tetratic order parameter evaluates to $T_\text{(2d)}=1$ when each pair of rods is either mutually parallel or perpendicular.

Similarly, in three dimensions, the discontinuities in the director field typically form grain boundaries, e.g, defect planes. 
To develop a classification concept in three dimensions, we construct in Appendix~\ref{app_tetratic} a 3d tetratic order parameter from the Steinhardt order parameters \cite{steinhardt1983bond}
\begin{equation}
I_l = \sum_{m=-l}^{l}|\braket{Y_{lm}}|^2
\label{invariant_maintext}
\end{equation}
with the spherical harmonics $Y_{lm}$.
Globally, this tetratic order parameter $T$ is given by
\begin{equation}
\begin{split}
\label{eq_tetr_methods_sec}
T = \frac{16\pi}{21 N^2} \Bigg| &\sum_{m=-4}^{4}\Bigg |\sum_{k=1}^{N}Y_{4m}(\theta_k,\phi_k)\Bigg|^2\\&
- \frac{3}{4}\sum_{m=-2}^
 {2}\left|\sum_{k=1}^{N}Y_{2m}(\theta_k,\phi_k)\right|^2\Bigg|. 
\end{split}
\end{equation}
This definition results in $T=0$ for an isotropic system, where the orientations $\{\uu_k\}$ are uniformly distributed on the unit sphere $S^2$ and $T=1$ for a system where all orientations are pairwise either parallel or orthogonal, i.e., if we have a local Cartesian coordinate system, where all rods are aligned to either of the axes. 
Analogously to the 2d tetratic order parameter $T_\text{(2d)}$, our definition \eqref{eq_tetr_methods_sec} of $T$ implies both perfect cubatic ($T=1,~S = 0$) and perfect nematic order ($T=1,~S = 1$) as special cases of $T=1$, such that we cannot measure this kind of tetratic order in a 3d system with either the standard cubatic~\cite{veerman1992phase,duncan2009theory} or the standard nematic order parameter. 

We now prove that the 3d tetratic order parameter~\eqref{eq_tetr_methods_sec} has the desired properties.
First, we show that it is 0 
for an isotropic system. In this case, the orientations $\{\theta_k,\phi_k\}$ approach a uniform distribution on the unit sphere $S^2$, such that the inner sums over $k$ in Eq.~\eqref{eq_tetr_methods_sec} approach an integral over $S^2$. This integral vanishes since the spherical harmonics satisfy
\begin{equation}
\INT{S^2}{}{\Omega}Y_{lm}(\theta,\phi)=0
\label{zero}
\end{equation}
for $l\neq 0$. Second, we show that it is 1 for a system where all particles are pairwise either parallel or orthogonal. Since $T$ is by construction invariant under coordinate transformations and since the functions $Y_{4m}$ and $Y_{2m}$ are invariant under parity transformation, we can assume without loss of generality that we have a configuration ($\Lambda$) of $a$ particles with orientation $(\theta,\phi)=(\pi/2,0)$, $b$ particles with orientation $(\theta,\phi)=(\pi,0)$ and $c$ particles with orientation $(\theta,\phi)=(\pi/2,\pi/2)$ (with $a,b,c \in \mathbb{N}_0$). The order parameter \eqref{eq_tetr_methods_sec} then evaluates to

\begin{equation}
\begin{split}
&T^{(\Lambda)} = \frac{16\pi}{21 (a+b+c)^2} \\&\cdot\Bigg| \sum_{m=-4}^{4}\left|a Y_{4m}\!\left(\frac{\pi}{2},0\right) + b Y_{4m}(\pi,0)+ c Y_{4m}\!\left(\frac{\pi}{2},\frac{\pi}{2}\right)\right|^2\\&
\ \ - \frac{3}{4}\sum_{m=-2}^
 {2}\left|a Y_{2m}\!\left(\frac{\pi}{2},0\right) + b Y_{2m}(\pi,0)+c Y_{2m}\!\left(\frac{\pi}{2},\frac{\pi}{2}\right)\right|^2\Bigg| \\ &=1, 
\end{split}\raisetag{3ex}
\label{eq_tetr_methods_sec1}%
\end{equation}
as can be easily confirmed by evaluating Eq.\ \eqref{eq_tetr_methods_sec1} using a computer algebra system.
Further examples are given in appendix \ref{app_examples}.

Finally, to generate a local field $T(\mathbf{r})$ of the tetratic order parameter from our simulation data, we sample the spherical harmonics entering Eq.~\eqref{eq_tetr_methods_sec} only within local spherical subsystems $B_{2.5D}(\mathbf{r})$,
which yields
\begin{equation}
\begin{split}
\label{eq_tetr_methods_sec_local}
T(\mathbf{r}) = \frac{16\pi}{21}\, \Bigg| &\sum_{m=-4}^{4}\left| \langle Y_{4m}(\theta_k,\phi_k) \rangle_{B_{2.5D}(\mathbf{r})}\right|^2\\&
- \frac{3}{4}\sum_{m=-2}^
 {2}\left|\langle Y_{2m}(\theta_k,\phi_k) \rangle_{B_{2.5D}(\mathbf{r})}\right|^2\Bigg| \,, 
\end{split}
\end{equation}
analogously to the nematic tensor, cf.~Eq.~\eqref{eq_Qsample}. 

\subsection{\label{sec_topcharge}Topological charge}
\begin{figure}[t!]
\begin{center}
\includegraphics[width=0.9\linewidth]{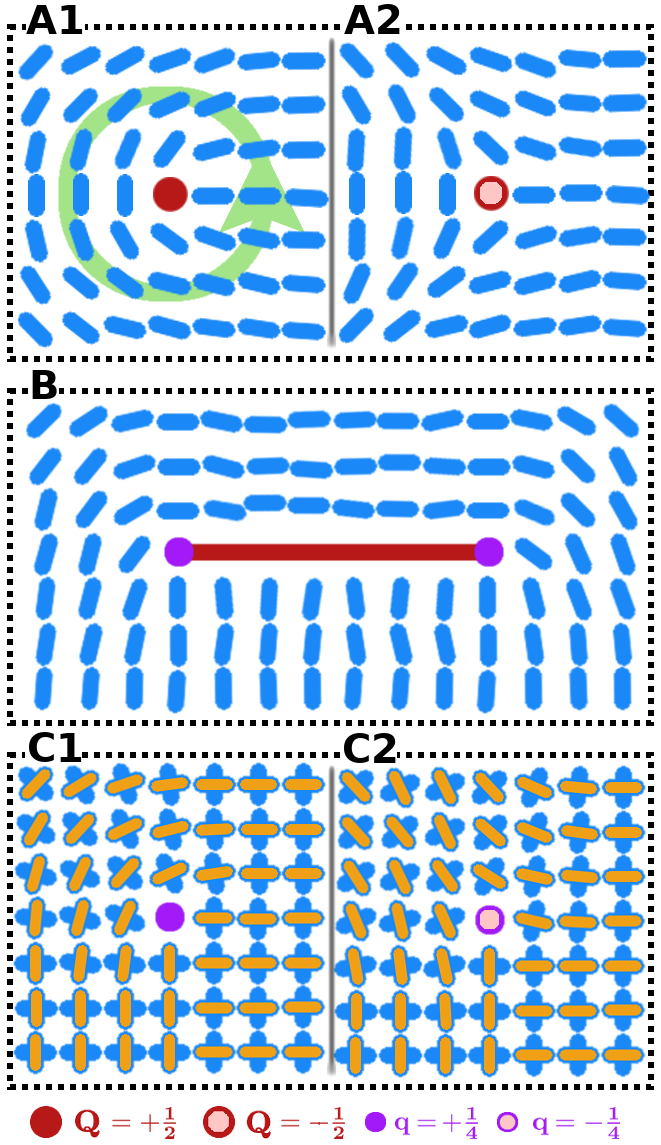}
\caption{\label{fig_topcharge}Schematic of the classification of 2d topological defects present as discontinuities in the director field $\nofr$. The topological charge of a defect corresponds to the net rotation of $\nofr$ traversing the defect in counter-clockwise direction (indicated by the arrow in \textbf{A1}) around the defects. \textbf{A:} Point defects for particles of $\pi$ rotational symmetry with charges $Q=+1/2$ (\textbf{A1}) and $Q = -1/2$ (\textbf{A2}) typically present in nematic liquid crystals. \textbf{B:} Decomposition of smectic grain boundaries into tetratic point defects.
Due to the preferred difference in orientation angles of $\pi/2$, the line defects can be classified as two isolated tetratic point defects of charge $q=\pm 1/4$. The schematic shows an exemplary line defect of total charge $Q=q_1+q_2=1/2$. \textbf{C:} Point defects for particles with $\pi/2$ rotational symmetry. Charges are $q = +1/4$ (\textbf{C1}) and $q = -1/4$ (\textbf{C2}). For ease of observing the continuous rotation, the two main axes are decorated differently.}
\end{center}
\end{figure}

Topological defects are identified as discontinuities in the director field $\nofr$, see Fig.~\ref{fig_topcharge}. In 2d nematic liquid crystals, the types of stable bulk defects are point defects. The strength of the defect, i.e., the degree of deformation of the surrounding fluid, is typically analyzed by a topological charge $Q$ that equates to the total rotation of the director traversing any contour $\mathcal{C}$ around the defect. This charge is given by the \textit{winding number} that can be explicitly calculated as the closed line integral along the contour $\mathcal{C}$ parametrized by $\kappa$, i.e.,
\begin{align}
Q = \frac{1}{2\pi} \oint_{\mathcal{C}(\mathrm{\kappa})}~\left [\hat{n}_1(\mathrm{\kappa}) \frac{\partial \hat{n}_2(\mathrm{\kappa})}{\partial \kappa}  - \hat{n}_2(\mathrm{\kappa}) \frac{\partial \hat{n}_1(\mathrm{\kappa})}{\partial \kappa} \right ] \mathrm{d\kappa},
\label{eq_Q}
\end{align}
where $\oint_{\mathcal{C}(\mathrm{\kappa})} \mathrm{d\kappa}=2\pi$ \cite{SoftMPhys}. 
Due to the apolarity of the particles, the configuration space of the orientations
is a semicircle with end points identified, commonly denoted by $S^1/Z_2$, i.e., we now consider $\mathbf{\hat{n}}$ to be a headless vector in the sense that we identify $\mathbf{\hat{n}}$ and $-\mathbf{\hat{n}}$ \cite{nem_defects}. 
From a topological point of view, we define the charge via the winding number because (since the winding number is a discrete quantity) different contours with different winding numbers can not be continuously transformed into each other, i.e., are not homotopic. All possible contours in the liquid crystal correspond to loops in $S^1/Z_2$. The fundamental group $\pi_1(S^1/Z_2)$ classifies loops in $S^1/Z_2$ up to homotopy equivalence. Consequently, all possible defects are classified by the fundamental group $\pi_1(S^1/Z_2)$. This group is given by $\frac{1}{2}\mathbb{Z}$ with the addition operation $+$. (The prefactor $1/2$ is a convention used in physics that is motivated by the geometric definition of charges via the winding number.) Therefore, the possible charges are $ Q\in\{k/2 | k \in \mathbb{Z}\}$ and these charges can be added to find the total charge of a combination of two defects. The sum of all charges is a conserved quantity in two dimensions (since it has to match the Euler characteristic of the confinement \cite{bowick2009two}). As an example, we illustrate $Q=1/2$ in Fig.~\ref{fig_topcharge}.\textbf{A1} and $Q=-1/2$ in Fig.~\ref{fig_topcharge}.\textbf{A2}. Moreover, the number of elements in $\pi_1(S^1/Z_2)$ is infinite, which is exemplified by the fact that the winding number can be any half-integer.

Defects typically observed in 3d nematics are disclination lines, along which the local orientational order is frustrated. (Point defects in three dimensions (\enquote{hedgehogs} \cite{nem_defects}) are not considered in this work.) In analogy to the 2d case, topological defects can be classified by considering closed loops in the orientational configuration space, up to homotopy equivalence. One typically analyses the defect in terms of the topology of a planar cross-section perpendicular to the disclination line \cite{frank1958liquid}. Since the configuration space of the orientations in three dimensions is a hemisphere with antipodal points identified, commonly denoted by $S^2/Z_2$, the rotation of $\nofr$ along $\mathcal{C}$ forms loops in $S^2/Z_2$, classified by $\pi_1(S^2/Z_2)$. By rotation into the third dimension, all half-integer disclinations can be continuously mapped onto each other. Correspondingly, $\pi_1(S^2/Z_2)$ has only two elements. As a result, for instance, 3d defects with cross-sections like in Fig.~\ref{fig_topcharge}.\textbf{A1} ($Q=+1/2$) and Fig.~\ref{fig_topcharge}.\textbf{A2} ($Q=-1/2$) are homotopically equivalent.
More specifically, the opposite charges $\pm 1/2$ correspond to opposite paths around half of the base of the hemisphere $S^2/Z_2$, both connecting two antipodal points. 
A $+1/2$ defect can be transformed into a $-1/2$ defect by passing the corresponding path in $S^2/Z_2$ over the north pole of the hemisphere \cite{nem_defects}.
This implies that (a) the charge defined by Eq.~\eqref{eq_Q} is no longer a conserved quantity and (b) it can no longer be used to classify the possible configurations of the nematic liquid crystal up to homotopy equivalence. There are only two topologically distinct configurations left, namely \enquote{defect} and \enquote{no defect}. (The discussion in this paragraph and the previous paragraph follows Ref.\ \cite{nem_defects}.)

\begin{figure}[t!]
\begin{center}
\includegraphics[height=1.05\linewidth]{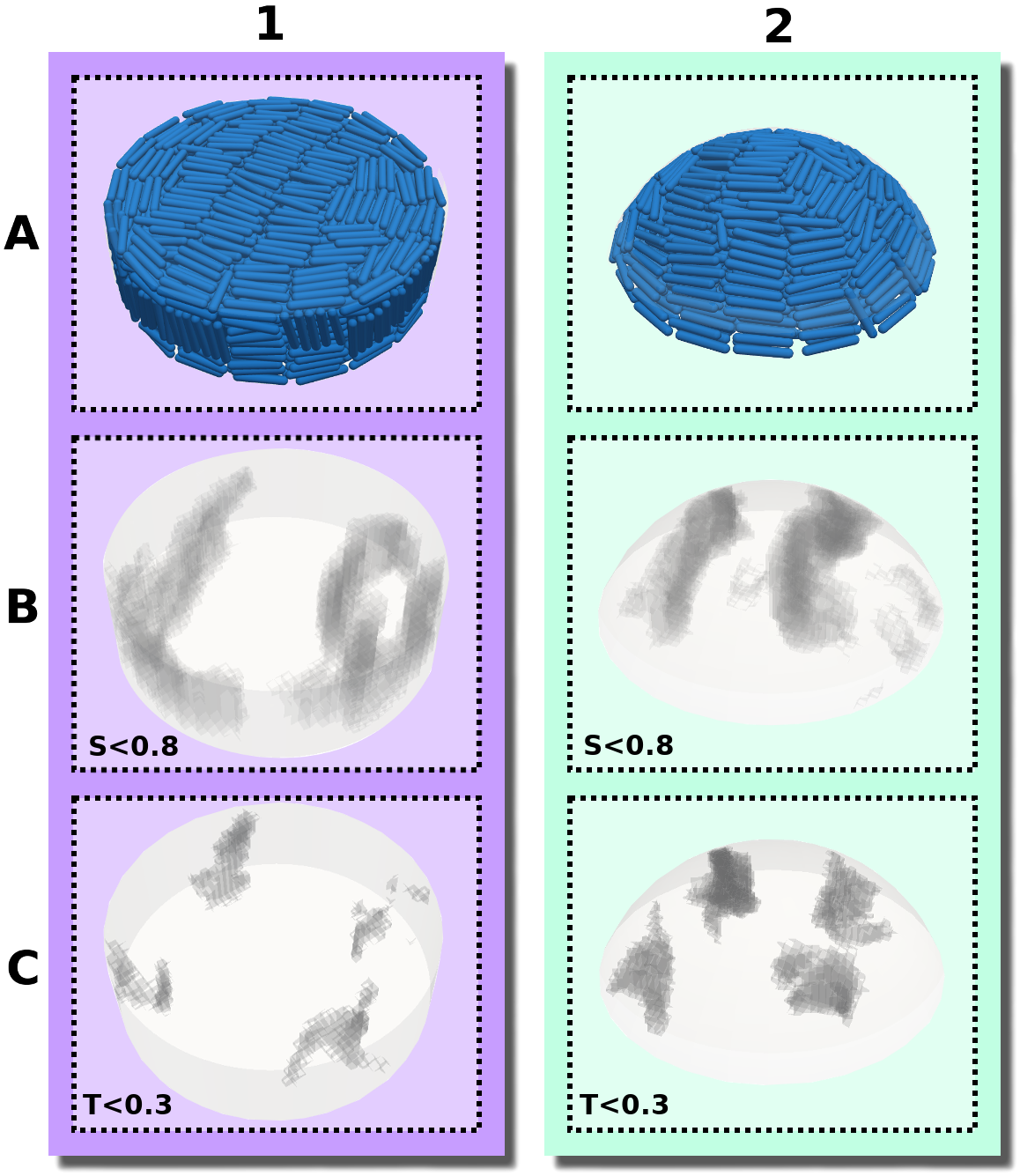}
\caption{Exemplary simulation results for two systems of 3d confined smectic liquid crystals. \textbf{Column 1:} Cylinder. \textbf{Column 2:} Spherical cap. Both systems feature height $h=2.4L$ and diameter $2R=8L$. \textbf{A:} Bird's eye view of the simulated system of hard rods. 
\textbf{B:} 3d visualization of the nematic defects according to the order parameter $\Sofr$. In order to observe the regions of low nematic order, i.e., defect regions, we display the data points with $\Sofr < 0.8$ as opaque gray clouds.
\textbf{C:} 3d visualization of the tetratic defects according to the 3d tetratic order parameter $\Tofr$. 
To visualize the defect regions, we only display the data points with $\Tofr < 0.3$. 
}
\label{fig_snapfig}
\end{center}
\end{figure}

Smectic liquid crystals, which additionally feature layering of the fluid particles, can be treated in the same spirit as nematics by
considering a vector field normal to the smectic layers \cite{machon2019,chen2009symmetry} or by directly working with the nematic director \cite{trebin1982topology,pindak1980macroscopic} (which coincides with the layer normals in the case of smectic-A order). However, if the smectic layers are sufficiently rigid, which is a prominent feature of colloidal systems, the discontinuities in the layered structure take the distinct form of elongated grain boundaries. Those grain boundaries are lines in two dimensions and planes in three dimensions.
Recent insight into the orientational topology of colloidal smectics in 2d \cite{monderkamp2021topology} suggests that these grain boundaries can be analyzed from the viewpoint of orientational topology by associating a topological charge to these defects as a whole. Furthermore, it has been shown that the rotation of the local director occurs mainly around the endpoints of the grain boundaries (see Fig.~\ref{fig_topcharge}.\textbf{B}). 
Those endpoints can be analyzed as isolated tetratic point defects by superimposing a tetratic director onto the fluid particles, i.e., considering orientations with $\pi/2$ rotational symmetry, where one of the axis points along the main axes of the rods (see Fig.~\ref{fig_topcharge}.\textbf{C}).
Due to the preferred difference of $\pi/2$ in the orientation angle across the grain boundary in smectics, those tetratic point defects display quarter charges $Q\in\{k/4 | k \in \mathbb{Z}\}$
(see Fig.~\ref{fig_topcharge}.\textbf{C}) \cite{monderkamp2021topology}. Geometrically speaking, this is a consequence of the fact that the rotation of the director around such a defect (divided by $2\pi$) is an integer multiple of $1/4$ (and not of $1/2$ as for standard nematic defects). Topologically speaking, this is a consequence of the fact that the tetratic order parameter superimposed in Fig.~\ref{fig_topcharge}.\textbf{C} takes values in $(S^1/Z_2)/Z_2$, which is a quarter-circle with end points identified. This order parameter becomes singular only at the endpoints, such that we can classify these endpoints as topological defects by integrating along a closed contour around them without having to pass through a singular point. The fundamental group is $\pi_1((S^1/Z_2)/Z_2)=\frac{1}{4}\mathbb{Z}$, where the conventional prefactor has now been set to $1/4$.

An important property of smectic structures is their rigidity due to the additional constraint provided by the positional order. As will be detailed in the results in \sect~\ref{sec_results}, the space occupied by the orientations $\{\uu_k\}$ is drastically reduced in our simulations of 3d colloidal smectic systems, i.e., all orientations are approximately perpendicular close to the line disclinations. Therefore, it is no longer possible to transform the defects with $Q=+1/2$ into defects with $Q=-1/2$, implying that they are topologically distinct and that the charge $Q$ defined by Eq.~\eqref{eq_Q} is effectively a conserved quantity. In this way, we construct a formalism for analyzing the 3d grain boundaries in \sect~\ref{sec_results} with the help of the previously defined 2d model \cite{monderkamp2021topology}.

\section{\label{sec_results}Results}
\subsection{\label{sec_res_OP}Detection of defects via order parameters}

Previous studies on 2d smectics in a simply-connected convex confining cavity
\cite{rectangularConfinement,garlea2016finite,slitpores,CollLQinSqConf,de2009topological}
have revealed the existence of a large, relatively defect\New{-}free central domain, encompassing several smectic layers, which connect opposite ends of the cavity. This bridge state can generally be observed for a large range of confinements \cite{monderkamp2021topology}. 
Indeed, we find that this benchmark structure also persists when extending the system into the third dimension.

\begin{figure}[t]
\begin{center}
\includegraphics[height=1.05\linewidth]{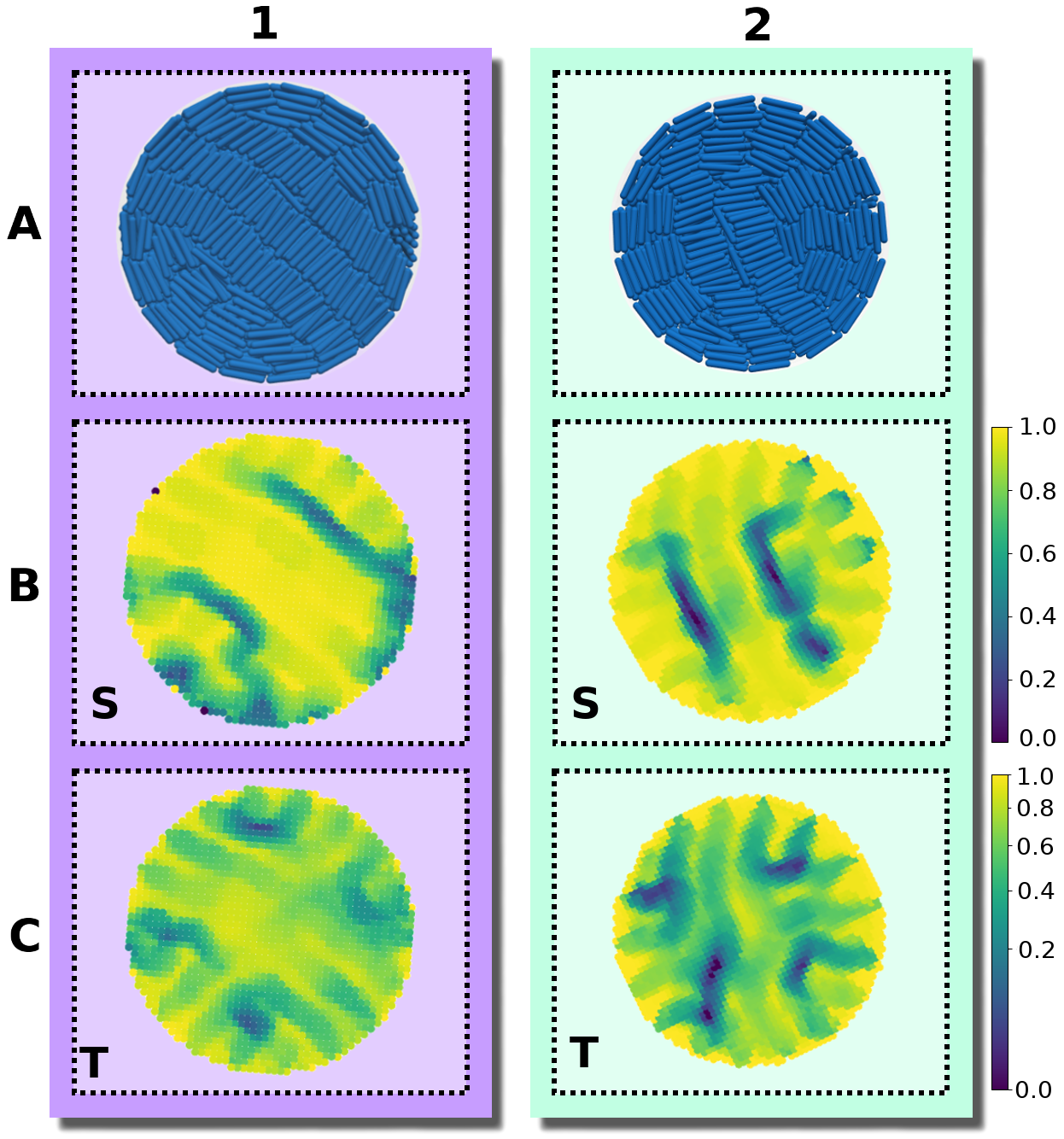}
\caption{\label{fig_snapfig2}Bottom view of the systems depicted in Fig.~\ref{fig_snapfig}. 
\textbf{Column 1:} Cylinder. \textbf{Column 2:} Spherical cap.
\textbf{A:} Bottom view of the snapshots to exemplify the structure of the 2d cross-sections. \textbf{B:} $\Sofr$ for the system projected onto the horizontal plane. \textbf{C:} $\Tofr$ for the system projected onto the horizontal plane.}
\end{center}
\end{figure}

In Figs.~\ref{fig_snapfig} and~\ref{fig_snapfig2}, we show typical simulation results, from a bird's eye view and in a 2d depiction, respectively, for two representative systems of hard rods confined to a cylindrical container and spherical cap. The snapshots in Figs.~\ref{fig_snapfig}.\textbf{A1} and \ref{fig_snapfig}.\textbf{A2} give an indication of the chosen dimensions of the confinement in terms of the rod length: The height $h$ of both cylinder and spherical cap is $2.4L$, while the diameter $2R$ of the footprint is fixed at $8L$. Both systems display what can be considered a generalized 3d bridge state.
This becomes even clearer when considering the bottom view of both systems in Figs.~\ref{fig_snapfig2}.\textbf{A1} and \ref{fig_snapfig2}.\textbf{A2}. Apparently, 
the bottom layer of rods in our 3d systems form 2d bridge states. 
Due to the symmetry of the cylinder, this structure is also mirrored on the top side. Even though the top surface of the spherical cap is strongly curved, the structure on it still resembles a 2d bridge state. 

To study the particle orientation throughout the system in more detail, we examine the topology of the corresponding order-parameter fields. In Figs.~\ref{fig_snapfig}.\textbf{B1} and \ref{fig_snapfig}.\textbf{B2}, we visualize the data points of low nematic order by showing the regions with $\Sofr<0.8$ in gray.
For both confinements, the resulting plots show a pair of planar disclinations, nestled to the sides of the central bridge domain.
These defect planes reach from the top to the bottom of the container, while their shape barely varies along the vertical axis. 
Moreover, at cross sections of constant height, they have very little contact to the outer walls. This picture is reinforced in Figs.~\ref{fig_snapfig2}.\textbf{B1} and \ref{fig_snapfig2}.\textbf{B2}, which show the nematic field of the systems projected on the horizontal plane, i.e., the plane perpendicular to the symmetry axis. Indeed, these projections closely resemble the nematic field $\Sofr$ for 2d systems, confirming that the shape of the defect planes varies little along the vertical axis. In addition to these grain-boundary planes, the simulated cylindrical structure gives rise to several spots where $\Sofr$ is significantly decreased at the mantle surface. As can be seen in Fig.~\ref{fig_snapfig}.\textbf{A1}, these spots correspond to locations where layers of single-rod depth align with the cylinder mantle.
For the spherical cap, the formation of such domains is largely suppressed by the curved boundary.

Figures~\ref{fig_snapfig}.\textbf{C1} and \ref{fig_snapfig}.\textbf{C2} similarly show regions of low 3d tetratic order according to the order parameter $\Tofr$. 
In analogy to the nematic case, we locate the defects by identifying the regions where the order is minimal.
Due to the relatively high sensitivity to orientational fluctuations of $\Tofr$, we display the data points only for $\Tofr<0.3$ in gray. It is then clearly visible that the nematic defect planes split into two tetratic defect pillars, each spanning from the bottom plane of the cavity to the top surface. Figures~\ref{fig_snapfig2}.\textbf{C1} and \ref{fig_snapfig2}.\textbf{C2} show the corresponding tetratic order-parameter field for the systems projected on the horizontal plane. This visualization shows that the minima in the order-parameter field are well localized and take an almost point-like shape. This again confirms that the tetratic disclinations in 3d appear as relatively straight lines, parallel to the vertical axis of the confinement.

In general, one identifies orientational defects as singular geometric objects in space, where the local director field $\nofr$ (see Eq.~\eqref{eq_Qsample}) jumps discontinuously. 
In particular, in 2d/3d liquid crystals with a smectic-A symmetry, the typical difference across any defect is $\pi/2$. As a result, nematic defects can be identified with the help of 
$\Sofr$. The same angular difference leads to a promotion of tetratic order everywhere except for the endpoints/edges, where the preferred orientation rotates. As a result, we identify a set of tetratic disclination points/lines, with the help of $\Tofr$, that sit on the endpoints/edges of each nematic defect.

\subsection{\label{sec_conf_sm_structure}Confined smectic structure} 

\begin{figure*}[t!]
\begin{center}
\includegraphics[width=0.85\linewidth]{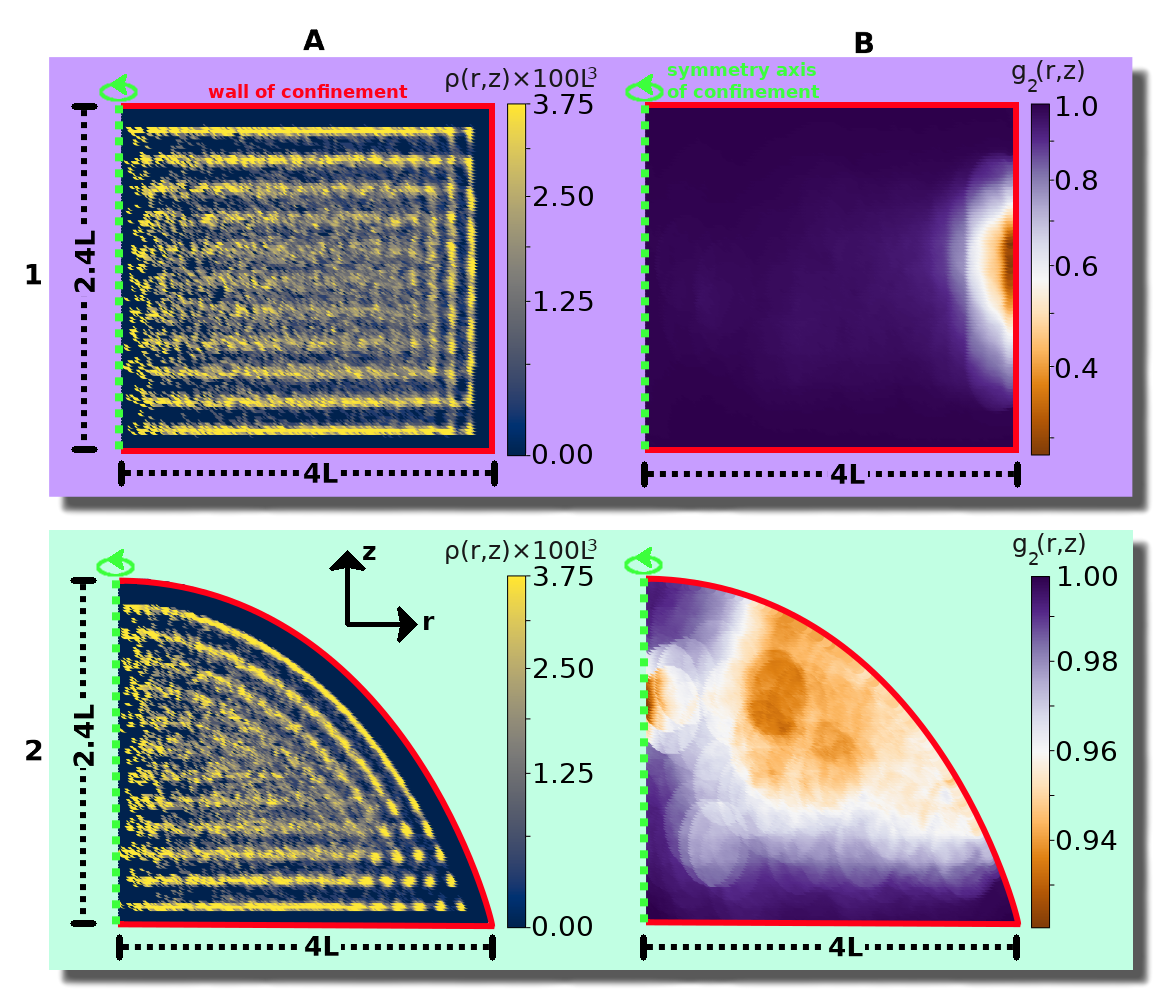}
\caption{\label{fig_bottom_correlation}Averaged structural properties of smectic liquid crystals confined in a cylinder (\textbf{row 1}) and spherical cap (\textbf{row 2}) of height $h = 2.4L$ and diameter $2R=8L$.
All diagrams are shown in cylindrical coordinates, averaged over the azimuth around the vertical center axis of the confinement (shown as a dashed green line). All diagrams are additionally averaged over 15 simulations. 
\textbf{A:} One-particle density $\ndens(r,z)$ (defined by Eq.~\eqref{eq_one_part_dens}). 
These fields exemplify the layered fluid structure along a radial slice of the confinement, influenced by the planar surface anchoring of the confinement. To improve the contrast, we set the interval of the color bar to $[0,3.75]$, ensuring that the structures are clearly visible. All values above $3.75$ are mapped to $3.75$. \textbf{B:} Correlation $g_2(r,z)$ (defined by Eq.~\eqref{eq_bottom_corr}) of the orientations at a certain position with the
orientations projected into the horizontal plane. This function illustrates the deviation from the preferred horizontal orientation depending on the position within the confinement.}
\end{center}
\end{figure*}

\begin{figure}[t!]
\begin{center}
\includegraphics[width=1.0\linewidth]{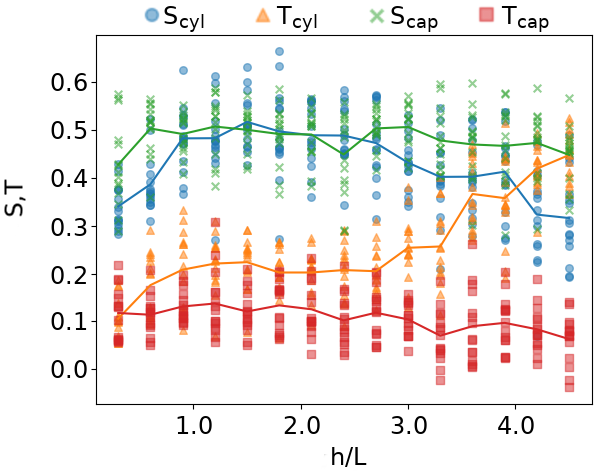}
\caption{\label{fig_TS_of_heights} Nematic order parameter $S$ and tetratic order parameter $T$ (see Eq.~\ref{sec_ordpar}) for the smectic liquid crystals
confined to 3d cavities in the shape of cylinders and spherical caps (see Fig.~\ref{fig_concept}). Along the horizontal 
axis we show a range of confinements with different heights $h$ in units of spherocylinder length $L$. The vertical 
axis represents different values of the order parameters evaluated for the entire system. For each confinement, we performed 15 simulations for each value of $h$. The height $h$ is swept from $h=0$ up to $h=4.5L$ in steps of $0.3L$. For each simulation run, we show two data points, corresponding to the two order parameters $S$ and $T$. The curves show the order parameters averaged over the 15 simulation runs.}
\end{center}
\end{figure}

To better understand the structural details of our confined smectics, we consider the number density 
\begin{equation}
\label{eq_one_part_dens}
\begin{split}
\!\!\!\!\!\!\!\!\ndens(r,z) = \frac{1}{NV_0} \sum^N_{k=1} \Theta\! \left( \frac{D}{10} -
\sqrt{(z-z_k)^2+ (r-r_k)^2} \right)\!\!\!\!\!
\end{split}
\end{equation}
of the center positions of the rods, averaged over the azimuth, where $r_k$ and $z_k$
refer to the positions of the particles in cylindrical coordinates.
Here, the Heaviside step function is denoted by $\Theta(x)$ and $V_0$ 
is the intersection volume of the respective toroidal bin with the container. We further divide by the particle number $N$, such that $\ndens(r,z)$ corresponds to the probability distribution for the position of a single particle.
Additionally, we show the local orientational distribution function 
\cite{zhao2007nematic,narayan2006nonequilibrium} 
\begin{equation}
g_2(r,z) = \langle P_2 (\sin\theta_k) \rangle_{B_{1.7D}(r,z)}, 
\label{eq_bottom_corr}%
\end{equation}
of the rods at position $(r,z)$, where $\sin\theta_k$ denotes the orientation of the $k$-th particle projected into the horizontal plane and $P_2(x)$ is the second Legendre polynomial. To obtain an appropriate resolution, we average within a spherical subsystem of radius $1.7D$.

Both quantities $\ndens(r,z)$ and $g_2(r,z)$ are averaged over 15 simulations runs, with different randomized initial states.
We show the resulting distribution functions in Fig.~\ref{fig_bottom_correlation} 
for the same confinements of container height $h=2.4L$ and diameter $2R=8L$, presented in \sect~\ref{sec_res_OP}. For ease of observation, the plots are stretched in the vertical direction. We generally observe for both confinements that the density profiles distinctly show the layering structure of the fluid, reflected by the relatively localized lines close to the outside walls. This indicates, that the positions of the layers are strongly influenced by the planar surface anchoring on the outer walls. 
While these peaks are less pronounced further inside the cavity, the diagrams show clear indication of horizontal stacking from the bottom to the top of the confinement. We stress that the kind of layering visible in the density profiles happens on the scale of the particle diameter $D$ and should not be confused with smectic layering along the direction of the rod axes of length $L+D$.

Along the vertical axis, the density profile for cylindrical confinement in Fig.~\ref{fig_bottom_correlation}.\textbf{A1} shows 11 layers of particles within a length interval of $2.4L=12D$, indicating that their typical orientation is horizontal.
This observation is reinforced in Fig.~\ref{fig_bottom_correlation}.\textbf{B1}, showing that the orientations of the rods strongly correlate with the horizontal plane in almost the whole container.
The depicted correlation function additionally indicates 
the presence of vertical rods close to the mantle of the cylinder, where $g_2(r,z)$ drops to $g_2\approx 0.3$, consistent with the occasional appearance of vertical rods on the perimeter of the cylinder shown in the snapshot in Fig.~\ref{fig_snapfig}.\textbf{A1}. 
Accordingly, the density profile in Fig.~\ref{fig_bottom_correlation}.\textbf{A1}
indicates a transition between vertical layers close to the mantle and horizontal layers in all other regions. In the corners, where the horizontal and vertical layers are compatible, we find fairly sharp isolated point-like peaks, exemplifying the high probability of a rod to sit aligned to both neighboring walls.

The density profile for the spherical-cap-shaped container in Fig.~\ref{fig_bottom_correlation}.\textbf{A2} indicates 12 stacked fluid layers in the middle of the confinement, indicating a stronger compression of the fluid than in the cylinder. Layers which are closer to the curved surface of the container are bent, while those closer to the bottom surface are horizontal. 
Again, we see localized peaks close to the corner,
which are even more pronounced than in the cylindrical container due to the smaller opening angle. 
Here, the roughly 20 isolated peaks are arranged on an approximately hexagonal grid, 
representing the structure of rods sitting at an angle of $\pi/2$ to both walls, at approximately the same distances to the perimeter in all simulations. This clearly demonstrates the influence of the extreme confinement. 
Figure~\ref{fig_bottom_correlation}.\textbf{B2} shows the orientational correlation of the rods $g_2$ with the bottom plane. It is visible that all rods are aligned fairly horizontally within the whole spherical cap (mind the different color scale compared to the cylindrical cavity). 
Only in the vicinity of the curved surface of the container, $g_2(r,z)$ is slightly reduced to values of $g_2\approx 0.9$, indicating that the rods rotate slightly out of the horizontal plane when aligning with the curved wall.

To study the effect of confinement height $h$ in more detail, we vary this geometrical parameter~from $h=0.3L$ to $h=4.5L$ in steps of $0.3L$, performing in each case 15 simulations for each confining geometry. In Fig.~\ref{fig_TS_of_heights}, we show the resulting nematic order parameter $S$ and the 3d tetratic order parameter $T$ evaluated for the entire system as a function of $h$. For the spherical cap, both order parameters $S$ and $T$ globally remain at fairly constant values, irrespective of the confinement height $h$. In detail, the nematic order parameter settles at $S\approx 0.5$, while the tetratic order parameter settles at $T\approx 0.1$, only showing a slight downwards trend. 
In stark contrast, for cylindrical confinement the tetratic order parameter increases strongly with increasing $h$, while the nematic order parameter displays a nonmonotonic behavior.

Comparing the ordering behavior in the two types of cavities in more detail,
we also observe in Fig.~\ref{fig_TS_of_heights} that for the most shallow confinements with $h=0.3L$
the two values of $T$ are qualitatively similar, whereas $S$ takes a slightly lower value in the cylinder than in the cap.
For this small height of the cavities, there are practically no effects of the third dimension, such that, like in a true 2d system, the aspect ratio $p=5$ of the rods considered here is too small to result in a significant orientational order and much less a smectic bridge state. 
The fact that the global nematic ordering in the spherical cap is still higher than in the cylinder, relates to the decreased accessible radius of the effective circular confinement.
Upon departing from this quasi-2d case by increasing $h$, the global order generally increases. For the cylinder, however, the nematic order parameter decreases again from $S\approx0.5$ to $S\approx0.3$, while $T$ drastically increases from  $T\approx0.1$ to $T\approx0.4$ as soon as $h>1.5L$. 
This behavior, which is specific for the cylindrical geometry,
can be explained by the fact that we observe a higher fraction of vertical rods on the mantle surface for taller confinements, reducing the global nematic order.  In turn, this alignment even leads to an increase of the tetratic order, since the vertical rods are perfectly perpendicular to the central domain. This kind of behavior is not observed for the spherical cap due to its curved boundary. We expect that, for even larger cylinder heights $h$, the majority of rods aligns with the mantle surface such that the global nematic order increases again.

\subsection{\label{sec_res_top_charge}Topological charge}
In \sect~\ref{sec_res_OP}, we discussed the detection of the topological defects with the help of the order parameters.
As elaborated in \sect~\ref{sec_topcharge}, topological defects that span between system boundaries, such as those in Figs.~\ref{fig_snapfig} and~\ref{fig_snapfig2}, can be assigned a topological charge defined via the net rotation of the director $\nofr$ along an encircling closed contour. 
More specifically, these contours can be defined within any cross section parallel to the bottom plane. Note that the sum of all topological charges, defined in this way, is a conserved quantity in 3d only under specific circumstances like in the smectic systems considered here, which can be understood as follows.
Imagine, for example, that Figs.~\ref{fig_topcharge}.\textbf{A1} and \ref{fig_topcharge}.\textbf{A2} show a cross section of a 3d nematic system. In this case, the $Q=+1/2$ defect from Fig.~\ref{fig_topcharge}.\textbf{A1} can be transformed into the $Q=-1/2$ defect from Fig.~\ref{fig_topcharge}.\textbf{A2} by flipping the orientations $\uu_k$ of all rods individually across the vertical picture axis. 
This transformation can be performed as a continuous mapping in 3d space, thereby obtaining a $Q=+1/2$ defect from a $Q=-1/2$ or vice versa.
Additionally, this rotation can occur continuously along a disclination line, resulting in different values for topological charges, depending on the respective chosen cross section \cite{afghah2018visualising}. 
As a result, all possible structures can in principle be either homotopically equivalent to a charge-free structure with $Q=0$ or to a defect structure with $Q=1/2$. 

In the previous Sec. \ref{sec_conf_sm_structure}, we elaborated that the confined smectic fluids in our simulations majorly consist of stacked quasi-2d layers.
By showing that the 3d systems consist of a number of stacked quasi-2d layers, with no out-of-plane rotation, the defects do not undergo the transitions mentioned above.
We thus argue that the topological charge is equal for all horizontal cross sections, such that we can consistently define charges of any of the defects visible in Fig.~\ref{fig_snapfig}. 
Additionally, we observe very similar structures on the top and bottom surface of the cylindrical systems and even on the curved surface of the spherical cap. This similarity of top and bottom structures is a further proof that the structure persists through all horizontal slices.
We can thus assume, to a good degree of approximation, that the director $\nofr$ does not rotate out of the 2d layers the 3d system consists of. This reduces the orientational configuration space from $S^2/Z_2$ to $S^1/Z_2$, implying that we can treat the topology in full analogy to the 2d case.
In this way, by observing the bottom plane, we can infer that the total topological charge of the nematic grain boundary is equal to $Q=1/2$ and matches the charge of the 2d counterpart. More specifically, the grain boundaries split into two pillars, i.e., tetratic disclination lines, with $q=1/4$ each, corresponding to $q=1/4$ point charges in the cross-sections.

Less frequent exceptions to the generic ordering behavior described above are given by the occasional alignment of the rods with the curved surface of the cap-shaped container as well as the small vertical clusters present at the mantle surface of the cylindrical container.
The former case does not undergo a transition between $\pm 1/2$ charged cross sections, as no rods are present that are drastically rotated out of plane. This is supported by the fact that the flat projection of the structure on the curved surface mirrors the structure of the bottom plane. In the cylindrical case, according to our observation, a director field $\nofr$ can most of the time be defined in a neighborhood around the defect, such that the vertical clusters do not influence the topological charges.

From mathematical topology it follows that for liquid crystals confined to 2d manifolds, the total topological charge in the director field has to match the Euler characteristic $\chi$ of the container \cite{bowick2009two}. The Euler characteristic $\chi$ is an algebraic invariant. Accordingly, results of previous work agree for nematic \cite{dzubiella2000topological} as well as smectic liquid crystals \cite{monderkamp2021topology} confined to simply-connected convex cavities and for smectics confined to 2d spherical surfaces embedded in 3d \cite{allahyarov2017}. In this work, we have presented 3d simply-connected convex confinements, where the sum of topological charges defined through integrating around a closed contour in a 2d cross section matches the Euler characteristic $\chi=1$ of this cross section.

\section{\label{sec_conclusion}Conclusion}
In this work, we provided an insight into the topology of defects in 3d smectic liquid crystals. In 3d smectic liquid crystals, orientational defects take the shape of extended planar grain boundaries, across which the local preferred direction jumps by an angle of $\pi/2$. Combining the established knowledge of classification of 3d nematic disclination lines with recent insights into the classification of grain boundaries as orientational defects in 2d smectic liquid crystals, we presented a formalism for the analysis of topological charge distribution. We exemplified this formalism on smectic structures in cylindrical and spherical cap containers, obtained using Monte-Carlo simulation.

In the 2d analysis one can utilize the coexistence of nematic and tetratic defects and, with the help of the tetratic order parameter, locate the points where the preferred direction rotates. Accordingly, we introduced a tetratic order parameter which can be readily applied to 3d systems. The 3d tetratic line defects were then analyzed along 2d cross sections. 

In a 3d system, the sum of the winding numbers of all defects is not in general a conserved quantity since defects with different winding numbers can be transformed into each other. However, we inferred from the rigidity of the smectic phase in our hard-rod system that the confined structures can be interpreted as stacked quasi-2d systems, such that topological charge behaves akin to electromagnetic charge and follows similar additive conservation laws.
We thus found in the simulated systems that the total topological charge matches the Euler characteristic $\chi=1$ of the containers and splits into two orientational defects, i.e., two grain boundaries with topological charge $Q=1/2$ each. Those can in turn be split into two tetratic disclination lines, with charges $q=1/4$. In general terms, we find it remarkable that the 2d topological charge, which does not have to be conserved for mathematical reasons in three dimensions, is conserved for physical reasons in the systems considered in this work.

Throughout this paper, we have in particular shed light on planar grain boundaries which split into two tetratic disclination lines.
 Our insight will thus be useful for the interpretation of future numerical~\cite{marechal2017density,chiappini2020}, experimental~\cite{CollLQinSqConf,annulus,lopezleon2012,repula2018} and theoretical~\cite{delasheras2006,wittmann2014,wittmann2016,marechal2017density,xia2021} research on confined smectic structures in three dimensions.
To this end, our topological picture can be extended to study more complex geometries and topologies in 3d, e.g., by observing $q=-1/4$ tetratic defects, which, in analogy to the 2d case, should emerge at junction points of defect networks in confinements that promote multiple domains \cite{monderkamp2021topology} or close to concave regions of the system boundary~\cite{annulus}. 
Of interest may also be the investigation of the connection of orientational defects to positional defects, such as dislocations and focal conic domains.
In nematic systems, it is widely accepted that the dynamical properties of a defect are influenced by the respective topological charge \cite{vromans2016orientational,harth2020topological,hydroOfDef,tan2019topological,crawford1996liquid,toptherdef}. 
Therefore, understanding the role of smectic orientational defects, i.e., grain boundaries analyzed as a connected pair of tetratic defects, in nonequilibrium is also of particular importance for understanding, e.g., nucleation processes \cite{maeda2003,schilling2004,ni2010,cuetos2010} and the dynamics in active smectics \cite{grelet2008dynamical,shankar2020topological}.

\section*{Acknowledgements}
We thank Arjun Yodh and Alice Rolf for helpful discussions. This work is funded by the Deutsche Forschungsgemeinschaft (DFG, German Research Foundation) -- VO 899/19-2; WI 4170/3-1; LO 418/20-2.
M.t.V.\ thanks  the Studienstiftung des deutschen Volkes for financial support. 

\appendix
\section{\label{app_tetratic}Tetratic order parameter in three dimensions}
In this appendix, we provide an appropriate definition of the tetratic order parameter to characterize the topological fine structure in smectic systems of uniaxial rods.

\subsection{Definition}
In general, a system of $N$ uniaxial particles can be microscopically described by a distribution function $f(\uu)$ that reads
\begin{equation}
f(\uu)=\frac{1}{N}\sum_{k=1}^{N}\delta(\theta-\theta_k)\delta(\phi-\phi_k),
\label{fmicro}
\end{equation}
where $\uu = (\sin\theta\cos\phi,\sin\theta\sin\phi,\cos\theta)^\mathrm{T}$ is the orientation vector.
Such a function can be expanded as \cite{te2020relations}
\begin{equation}
f(\uu)=\sum_{l=0}^{\infty} \sum_{i_{1},\dotsc,i_{l}=1}^{3} \ff{3}{l}{i_{1}\dotsb i_{l}} u_{i_{1}}\!\dotsb u_{i_{l}},
\label{fu}%
\end{equation}
where 
$u_{i}$ is the $i$-th element of the orientation vector and the expansion coefficients are given by \cite{te2020relations}
\begin{equation}
\ff{3}{l}{i_{1}\dotsb i_{l}} = \frac{2l+1}{4\pi} \INT{S^{2}}{}{\Omega} f(\uu) \TT{3}{l}{i_{1}\dotsb i_{l}}
\label{eq:fdD}
\end{equation}
with the tensor Legendre polynomials $\TT{3}{l}{i_{1}\dotsb i_{l}}$. An expansion of the form \eqref{fu} is also possible in a 2d system, in this case $\uu$ is a 2d vector depending on just one angle and the expression \eqref{eq:fdD} is slightly modified (see Ref.\ \cite{te2020relations}). The second-order contribution $\ff{3}{2}{i_1 i_2}$ is the nematic tensor. (In the main text (\cref{eq_Qsample}), we have, as is common, defined it with a different normalization that corresponds to multiplying the one resulting from \cref{eq:fdD} by $4\pi/5$.) An interesting mathematical property of the Cartesian expansion \eqref{fu} is that it is orderwise equivalent to the spherical multipole expansion 

\begin{equation}
f(\theta,\phi)=\sum_{l=0}^{\infty}\sum_{m=-l}^{l} f_{lm} Y_{lm}(\theta,\phi)
\label{fthetaphi}%
\end{equation}
with the spherical harmonics $Y_{lm}$ and the expansion coefficients
\begin{equation}
f_{lm}=\INTOII f(\theta,\phi) Y^\star_{lm}(\theta,\phi),
\label{flm}%
\end{equation}
where $\star$ denotes a complex conjugation \cite{te2020relations,te2020orientational}.

We are now looking for an order parameter that identifies configurations as ordered if the rods are either parallel or orthogonal to each other. In the 2d case, this can be simply done by superimposing tetratic order \cite{monderkamp2021topology}, i.e., fourfold rotational symmetry.
Mathematically, this corresponds to measuring defects not in the nematic order-parameter field, corresponding to the second-order term in the 2d version of Eq.~\eqref{fu}, but in the fourth-order contribution. 
This suggests that the desired order parameter can be constructed from the fourth-order term $(l=4)$ in Eq.~\eqref{fu} also in the 3d case. Since the Cartesian order parameter at fourth order has 81 components of which, due to symmetry and tracelessness, only 9 are independent, it is more convenient to work with the expansion coefficients of the angular expansion \eqref{fthetaphi} instead. 

From Eqs.~\eqref{fmicro} and \eqref{flm} we would then get the order parameters
\begin{equation}
f_{4m}= \frac{1}{N}\sum_{k=1}^{N}Y^\star_{4m}(\theta_k,\phi_k)
\label{flmdelta}
\end{equation}
of order $l=4$.
The values of these order parameters \eqref{flmdelta} depend, however, on the choice of the coordinate system. We now make use of the fact that the quantity
\begin{equation}
I_l = \sum_{m=-l}^{l}|\braket{Y_{lm}}|^2
\label{invariant}
\end{equation}
with the average $\braket{Y_{lm}}$ is an invariant of the spherical harmonics \cite{steinhardt1983bond} (i.e., it takes the same value for all choices of the coordinate system). This fact has been exploited in the study of bonding in liquids \cite{steinhardt1983bond,lechner2008accurate,tanaka2012bond} or orientational order in liquid crystals \cite{john2008phase}. Consequently, we should consider $I_4$ instead of $f_{4m}$.

Finally, we also need to take into account that the order parameter constructed from the invariants $I_l$ should (a) be normalized -- this can simply be ensured by multiplying it by an appropriate prefactor -- and (b) not distinguish between parallel and orthogonal rods.
Unfortunately, the invariant $I_4$ gives a larger value for parallel than for orthogonal configurations. To correct for this, we exploit the fact that the invariant $I_2$ measures nematic order \cite{john2008phase,blaak1999cylinders}, such that it is large for parallel configurations. 
Hence, our generalized order parameter should be proportional to $I_4 - \beta I_2$, where $\beta$ is a suitable prefactor. We have found an appropriate choice to be $\beta = 3/4$. 
Thus, we arrive at the tetratic order parameter
\begin{equation}
\begin{split}
T = \frac{16\pi}{21 N^2} \Bigg| &\sum_{m=-4}^{4}\left|\sum_{k=1}^{N}Y_{4m}(\theta_k,\phi_k)\right|^2\\&
- \frac{3}{4}\sum_{m=-2}^
 {2}\left|\sum_{k=1}^{N}Y_{2m}(\theta_k,\phi_k)\right|^2\Bigg| 
\end{split}
\label{t4}
\end{equation}
stated in Eq.~\eqref{eq_tetr_methods_sec}. The prefactor ensures a proper normalization.
Moreover, we use absolute values to ensure that $T$ is always positive. 
We have tested a range of possible configurations and found that $I_4 - \beta I_2 < 0$ is measured only for isotropic systems with $\left | I_4 - \beta I_2 \right | \ll 1$ (probably due to numerical fluctuations). This is reinforced by the notion that $I_2$ measures nematic order and systems with high nematic order result in $T \approx 1$.

\vspace{0.6cm}

\subsection{Examples \label{app_examples}}
To get an exemplary system that should be perfectly ordered by our definition, consider three orthogonal particles with orientations $(\theta_1,\phi_1)=(\pi/2,0)$, $(\theta_2,\phi_2)=(\pi,0)$ and $(\theta_3,\phi_3)=(\pi/2,\pi/2)$. We find 
\begin{equation}
\begin{split}
T =& \frac{16\pi}{21\cdot3^2}\Bigg|\sum_{m=-4}^{4}\left|Y_{4m}\left(\frac{\pi}{2},0\right) + Y_{4m}(\pi,0)+Y_{4m}\left(\frac{\pi}{2},\frac{\pi}{2}\right)\right|^2 \\
&- \frac{3}{4}\sum_{m=-2}^{2}\left|Y_{2m}\left(\frac{\pi}{2},0\right)+Y_{2m}(\pi,0)+Y_{2m}\left(\frac{\pi}{2},\frac{\pi}{2}\right)\right|^2\Bigg|\\=&1\,,\qquad
\end{split}\raisetag{3ex}
\end{equation}
as required. 
Similarly, for three parallel particles with orientations $(\theta_1,\phi_1)=(\theta_2,\phi_2)=(\theta_3,\phi_3)=(\pi/2,0)$, we get
\begin{equation}
\begin{split}
T = \frac{16\pi}{21\cdot3^2}\Bigg|&\sum_{m=-4}^{4}\left|3 Y_{4m}\left(\frac{\pi}{2},0\right)\right|^2\\&
- \frac{3}{4}\sum_{m=-2}^{2}\left|3 Y_{2m}\left(\frac{\pi}{2},0\right)\right|^2\Bigg|\\=1\,.\qquad&
\end{split}
\end{equation}
Finally, we flip the orientation of one particle by $\pi$ to show that this leaves the order parameter invariant, implying that it is apolar. We find
\begin{equation}
\begin{split}
T = \frac{16\pi}{21\cdot3^2}\Bigg|&\sum_{m=-4}^{4}\left|2 Y_{4m}\left(\frac{\pi}{2},0\right)+Y_{4m}\left(\frac{\pi}{2},\pi\right)\right|^2\\
&- \frac{3}{4}\sum_{m=-2}^{2}\left|2 Y_{2m}\left(\frac{\pi}{2},0\right)+Y_{2m}\left(\frac{\pi}{2},\pi\right)\right|^2\Bigg|\\=1\,,\qquad&
\end{split}\raisetag{3ex}
\end{equation}
such that Eq.~\eqref{t4} constitutes a solid basis for exploring tetratic order phenomena in three dimensions.

\vspace{0.3cm}


\end{document}